\documentclass[10pt,twocolumn,amssymb,amsmath,nobibnotes,nofootinbib,aps,prb,showpacs,floatfix]{revtex4-2} 
\usepackage{graphicx}
\usepackage{microtype}
\usepackage{multirow}
\usepackage{bm}
\usepackage{gensymb}
\usepackage{dcolumn}
\usepackage[dvipsnames]{xcolor}
\usepackage{graphicx}
\usepackage{dcolumn}
\usepackage{bm}
\usepackage{soul}
\usepackage{float}
\usepackage[hidelinks=true]{hyperref}
\usepackage[hyphenbreaks]{breakurl}
\hypersetup{
  colorlinks   = true, 
  urlcolor     = blue, 
  linkcolor    = blue, 
  citecolor    = blue 
}
\begin{document}

\title{Suppression of Intrinsic Hall Effect through Competing Berry Curvature in Cr$_{1+\delta}$Te$_2$}






\author{Prasanta Chowdhury$^1$, Jyotirmay Sau$^2$, Mohamad Numan$^1$, Jhuma Sannigrahi$^3$, Matthias Gutmann$^4$, Saurav Giri$^1$, Manoranjan Kumar$^2$, Subham Majumdar$^1$}
\email{sspsm2@iacs.res.in}

\affiliation{$^1$School of Physical Sciences, Indian Association for the Cultivation of Science, 2A \& B Raja S. C. Mullick Road, Jadavpur, Kolkata 700 032, India}

\affiliation{$^2$Department of Condensed Matter and Materials Physics, S. N. Bose National Centre for Basic Sciences, JD Block, Sector III, Salt Lake, Kolkata 700106, India}

\affiliation{$^3$School of Physical Sciences, Indian Institute of Technology Goa, Farmagudi, Goa 403401, India}

\affiliation{$^4$ISIS Neutron and Muon Source, Science and Technology Facilities Council, Rutherford Appleton Laboratory, Chilton Didcot OX11 0QX, United Kingdom}

\begin{abstract}
We conducted a comprehensive analysis of the magnetic and electronic transport properties of the layered chalcogenide Cr$_{1+\delta}$Te$_2$ in its single crystalline form. This material exhibits a ferromagnetic transition at a critical temperature of $T_C = 191$ K, characterized by significant thermal hysteresis in the magnetization data below this temperature. Measurements of isothermal magnetization, magnetocaloric effect, and magnetoresistance indicate that the system exhibits strong magnetocrystalline anisotropy, with the $c$-axis serving as the easy axis of magnetization. The Cr$_{1+\delta}$Te$_2$ compound shows  pronounced anomalous Hall effect (AHE); however, existing experimental and theoretical data do not provide a clear understanding of the nature and origin of this phenomenon. Our experimental findings suggest that the skew scattering mechanism primarily accounts for the observed AHE. In contrast, our theoretical study reveals the presence of gapped nodal points accompanied by non-zero Berry Curvature, which are expected to contribute towards intrinsic AHE. A detailed analysis of the electronic band structure, obtained through density functional theory calculations, reveals that the Berry Curvature  at different nodal points exhibit both positive and negative signs. These opposing contributions largely cancel each other out, thereby significantly diminishing the intrinsic contribution to the AHE.

\end{abstract}

\maketitle

\section{Introduction}

The discovery of intrinsic long-range magnetic ordering in two dimensional (2D) van der Waals (vdW) materials such as  Cr$_2$Ge$_2$Te$_6$~\cite{CrGeTe}, CrI$_3$~\cite{CrI3}, Fe$_n$GeTe$_2$ ($n$ = 3,4,5)~\cite{Fe3GeTe,Fe4GeTe,Fe5GeTe,sau} has  garnered significant research attention due to their potential applications in spintronics and magnetic storage devices~\cite{application1,application2,application3}. According to the Mermin-Wagner theorem~\cite{Wagner}, a 2D isotropic  Heisenberg type spin system can not achieve a long-range magnetic ordered state at non-zero temperature because of strong thermal fluctuations. However, the presence of strong magnetocrystalline anisotropy, often found in low-dimensional materials, can stabilize long-range order within the system~\cite{CrI3}.

\par
Among all the vdW materials, transition metal dichalcogenides (TMDs) attracted considerable research interest due to their structural flexibility, allowing for the intercalation of atoms between vdW layers and enabling tunable magnetic properties. In particular, magnetic TMDs drew attention for their potential to host non-collinear spin textures, such as chiral helimagnetism in Cr$_{1/3}$NbS$_2$~\cite{NonCo1}, skyrmion bubbles in Fe$_3$GeTe$_2$~\cite{SkBubble_Fe3GeTe2} \emph{etc}. Among magnetic TMDs, CrTe$_2$ stands out for exhibiting long-range magnetic order even in monolayer form \cite{CrTe2(2021),CrTe2(2022)}. One special aspect of these layered compounds is their ability to accommodate additional Cr atoms within the vdW gaps of CrTe$_2$ across a broad concentration range. This flexibility allows tunable Curie temperature, magnetism and structural properties~\cite{CrTe(PRM)}. 

Depending on the synthesis conditions and intercalated Cr concentration, a variety of Cr$_{1+\delta}$Te$_2$ compounds have been reported, such as CrTe$_2$ ($\delta$=0)~\cite{CrTe2(2021)}, Cr$_5$Te$_8$ ($\delta$=0.25)~\cite{trig_Cr5Te8}, Cr$_2$Te$_3$ ($\delta$=1/3)~\cite{Cr2Te3_Nanoscale}, Cr$_3$Te$_4$ ($\delta$=0.5)~\cite{Cr3Te4_InChem}, and CrTe ($\delta$=1)~\cite{CrTe_Jal.Com}. All these compounds are formed by alternative staking of Cr-rich (CrTe$_2$ layer) and Cr-deficient (intercalated Cr layer) layers along the  $c$-axis~\cite{Jpcm_structure}. Compounds such as Cr$_5$Te$_8$, Cr$_2$Te$_3$, and Cr$_3$Te$_4$ have been found to crystallize in monoclinic or trigonal structures. At higher Cr concentration, the compositions tends to adopt a NiAs type hexagonal structure, as observed in Cr$_5$Te$_6$~\cite{Cr5Te6_jmmm} and Cr$_7$Te$_8$~\cite{Cr7Te8_jpsj}. Some of these Cr-Te based magnetic compounds have been reported to exhibit anomalous Hall effect (AHE), which originates from either intrinsic or extrinsic mechanisms.

\par
The development of AHE in magnetic materials is one of the most intriguing aspects of condensed matter physics. AHE arises from three primary mechanisms: skew scattering, side-jump, and intrinsic deflection~\cite{Nagaosa,Yue2017}. Karplus and Luttinger first proposed a model related to the band structure of ferromagnetic metals incorporating spin-orbit interaction (the intrinsic KL mechanism)~\cite{karplus}, demonstrating that the anomalous Hall resistivity ($\rho^{AHE}_{xy}$) scales quadratically with the longitudinal resistivity ($\rho_{xx}$). Later, J. Smit~\cite{SMIT21,SMIT24} and Berger~\cite{Berger} identified two fundamental extrinsic mechanisms: skew-scattering and side-jump. Akin to intrinsic KL mechanism, side-jump scattering exhibits a quadratic relationship with $\rho_{xx}$, while skew-scattering has a linear relationship with $\rho_{xx}$. The inherent KL mechanism is directly connected to the Berry curvature (BC) of the occupied electronic Bloch states~\cite{Jungwirth,nagaosa2006anomalous,Gradhand_2012,PhysRevB.104.195108}. In recent decades, it has been realised that the intrinsic contribution of AHE is often governed by the non-vanishing BC~\cite{Kubler2014,sau2024hydrostatic,PhysRevB.107.125138}, rather than being directly tied to magnetization. By manipulating a material's electronic band structures and symmetries, it is possible to tune the BC and the inherent AHE~\cite{Shindou}. 

\par
Interestingly, recent experimental observations indicate that the electronic properties of Cr-Te compounds are sensitive to their composition~\cite{Fujisawa,AHE_trig_Cr5Te8,AHE_mono_Cr3Te4,Adv.Mat}. In addition to the AHE, several members of this series exhibit the topological Hall effect (THE), which is associated with either non-collinear spin structures, as seen in Cr$_5$Te$_8$~\cite{trig_Cr5Te8}, Cr$_3$Te$_4$~\cite{AHE_mono_Cr3Te4}, or skyrmionic spin texture, observed in Cr$_{1+\delta}$Te$_2$ ($\delta\sim$ 0.33)~\cite{NatCom_RanaSaha} and Cr$_{1.53}$Te$_2$~\cite{Adv.Mat}. Despite considerable studies on Cr-Te based quasi-2D compounds, the mechanism behind AHE remains elusive, and it is essential to shed more light on the nature and origin of the Hall effect in Cr$_{1+\delta}$Te$_2$. We have conducted a joint experimental and theoretical investigation on the composition Cr$_{1.33}$Te$_2$ belonging to the series, which has been reported to show THE~\cite{NatCom_RanaSaha}. In this work, we have experimentally investigated the structural and anisotropic magneto-transport properties of the compound. Our Hall data indicate that the AHE arises from the skew scattering mechanism, and we fail to observe the signature of THE in our studied single crystalline sample of Cr$_{1.33}$Te$_2$. Our theoretical band structure calculations reveal the presence of non-vanishing BC; however, their competing nature suppresses the overall intrinsic contribution to the AHE.

\begin{figure*}
	\centering
	\includegraphics[width = 16 cm]{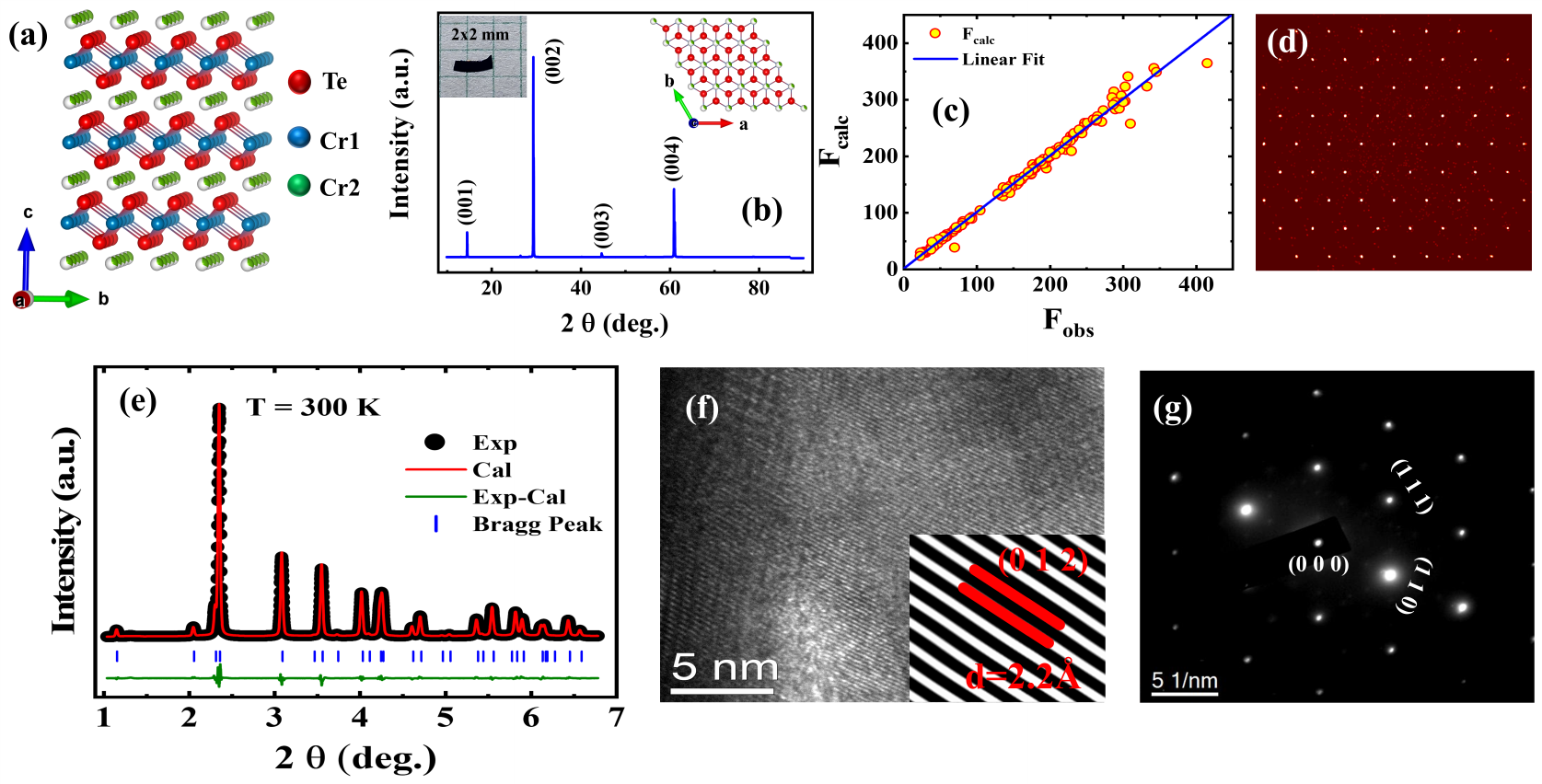}
	\caption{ (a) Crystal structure of Cr$_{1+\delta}$Te$_2$. (b) XRD pattern of single crystalline Cr$_{1+\delta}$Te$_2$ showing $(0~0~l)$ diffraction peaks. (c) Observed (F$_{obs}$) vs calculated (F$_{calc}$) structure factors of SCXRD data. (d) SCXRD reciprocal space precession image of $(h~k~2)$ layer. (e) Powder XRD pattern (symbols) with Rietveld refinement (solid line) at 300 K. (f) Enlarged HRTEM image displaying $(0~1~2)$ plane. (g) SAED pattern displaying different planes. }
	\label{fig:XRD}
\end{figure*}

\section{METHODS}
\subsection{Experimental Techniques}

High-quality single crystals of Cr$_{1+\delta}$Te$_2$ ($\delta$=0.33) were synthesized by chemical vapor transport (CVT) method~\cite{Adv.Mat}. High purity powders of elemental Cr (99.95 \% Alfa Aesar) and Te (99.997 \% Sigma Aldrich) were mixed at a gradient molar ratio and sealed in an evacuated quartz tube, with a small amount of iodine as a transport agent. The quartz tube is placed in a two-zone furnace for one week with a temperature gradient of 950-850 \degree C. The quartz tube was then cooled to room temperature, and the obtained plate like crystals were washed with acetone to remove excess iodine on the surface.
\par
Crystal structure and phase purity of the sample were investigated through single crystal x-ray diffraction (SCXRD) and powder x-ray diffraction (PXRD). SCXRD data were collected using a Rigaku-Oxford diffraction Xtalab synergy single-crystal diffractometer equipped with a HyPix hybrid pixel array detector using Mo-K$\alpha$ radiation. The SCXRD data reduction and reciprocal space precision image construction was done using CrysAlisPro software. Crystal structure was solved using SHELXT program package~\cite{SHELX} and refinement was done using Jana2020 software~\cite{jana2020}. High-resolution (wavelength $\lambda$=0.122 \AA ) temperature dependent PXRD data was collected from P21.1 beamline of PETRA III, DESY, Germany, recorded in the range of 10-300 K. For structural analysis, the Rietveld refinement of the PXRD data was performed using FULLPROF software~\cite{fullprof}. Compositional analysis of the single crystals was done through energy dispersive x-ray spectroscopy (EDXS) measurement using JEOL JSM-6010LA scanning electron microscope (SEM). High resolution transmission electron microscopy (HRTEM) and selected area electron diffraction (SAED) were performed using a JEOL TEM 2010. 
\par
Magnetic measurements were carried out using the vibrating sample magnetometer module of a commercial physical properties measurement system (PPMS, Quantum Design) as well as on a SQUID-VSM (MPM3) of Quantum Design. Hall and magneto-resistance (MR) measurements were performed on a cryogen-free high magnetic field system (Cryogenic Ltd. UK).
\par
\subsection{ Theoretical Calculations }
We used the VASP package~\cite{hafner2008ab} with a plane-wave basis set and pseudopotentials to analyse the electronic band structure of actual materials by \emph{ab} initio DFT calculations. The cutoff energy for plane waves is 600 eV, and the exchange-correlation potential is represented by the generalized-gradient approximation (GGA+U)~\cite{perdew}. The k-space integrations were carried out on an 8 $\times$ 8 $\times$ 8 grid. All structures were relaxed until the forces were less than 0.001 eV/Å. Matching the magnetic moments requires an effective Coulomb interaction (U) in the range of 0.5-0.8 eV. We used the WANNIER90 to construct Wannier functions from the DFT band structure~\cite{pizzi2020wannier90,MarzariPhysRevB.56.12847}.


\section{Results and Discussion}
\subsection{Sample Characterizations}

From EDXS measurements, the actual chemical composition of the compound is found to be Cr:Te = 1.33(1):2. X-ray diffraction taken from the flat surface of the plate-like Cr$_{1+\delta}$Te$_2$ single crystal shows the presence of only sharp $(0~0~l)$ peaks [Fig.~\ref{fig:XRD} (b)], indicating that the $c$-axis is perpendicular to the large flat surface of the crystal. From room temperature SCXRD refinement, our Cr$_{1+\delta}$Te$_2$ sample is found to crystallize in a trigonal crystal structure belonging to the space group \textit{P}$\overline{3}$m1 (164). The good linear behavior between the observed ($F_{obs}$) versus calculated ($F_{calc}$) structure factor plot [shown in Fig.~\ref{fig:XRD} (c)] signifies the refinement is quite satisfactory. The refined parameters are given in table~\ref{table:xrd}. Fig.~\ref{fig:XRD} (d) shows the reciprocal space precession image of $(h~k~2)$ layer and it shows a hexagonal lattice in this layer with well-defined spots. Fig.~\ref{fig:XRD} (a) shows the crystal structure of Cr$_{1+\delta}$Te$_2$, here the excess Cr atoms (Cr2) are intercalated in the vdW gaps created by two CrTe$_2$ layers. The Rietveld refinement of the PXRD data at 300 K [shown in Fig.~\ref{fig:XRD} (e)] with \textit{P}$\overline{3}$m1 space group also confirms the phase purity and the crystal structure of the sample. The refined lattice parameters and atomic coordinates are summarized in table~\ref{table:xrd}. The lattice parameters are pretty similar to the values reported for Cr$_{1.33}$Te$_2$~\cite{NatCom_RanaSaha} indicating that the Cr concentration of our single crystal sample is indeed close to 1.33. Additionally, the magnetic transition temperature and the temperature evolution of magnetization of our sample  match quite well with the previously reported Cr$_{1.33}$Te$_2$ sample. The HRTEM image shown in Fig.~\ref{fig:XRD} (f) further confirms the undistorted plane or absence of disorder in the crystal. The inset of Fig.~\ref{fig:XRD} (f) shows the spacing between the plane to be 2.2 \AA, which matches well with the inter-planar spacing $d_{(012)}$ of (0~1~2) plane as obtained from the x-ray diffraction studies. Fig.~\ref{fig:XRD} (g) shows the SAED pattern with the assigned miller indices.


\begin{table}
    \caption{Refined structural parameters of our Cr$_{1+\delta}$Te$_2$ sample obtain from SCXRD aand PXRD at $T$=300 K. }
	\label{table:xrd}
	\centering
    \setlength{\tabcolsep}{7.5pt}
	\begin{tabular}{c   c   c   c   c  c}
		\hline
        \hline
		Method &  & SCXRD &  & PXRD &   \\
        Temperature &  &300 K  & &300 K  &   \\
        Chemical Formula & &Cr$_{1.33}$Te$_2$  &  &Cr$_{1.33}$Te$_2$ & \\
        Space Group &  & \textit{P}$\overline{3}$m1  &  & \textit{P}$\overline{3}$m1 &  \\
        a(\AA)  &  & 3.93094(16)  & & 3.92981(7) &  \\
        c(\AA)  &  & 6.0522(3)  &  & 6.05663(13) &  \\
        V(\AA$^3$) &  & 80.991(6) &   & 81.003(3) &  \\
        $R$,$wR2$/$R_{exp}$,$R_{wp}$ &  & 3.28, 7.65 &   & 4.32, 6.12  &  \\
        $GOF$/$\chi^2$ &  & 2.03 &   & 2.01 &  \\
        \hline
	\end{tabular}
    \begin{tabular}{c   c   c   c   c  c}    
        Atom &   Wyck.  & x & y & z & Occ.   \\
        \hline
        Cr1 &   1b  & 0 & 0 & 1/2 & 1.0   \\
        Cr2 &   1a  & 0 & 0 & 0 & 0.33(1)   \\
        Te &   2d  & 1/3 & 2/3 & 0.2489(2) & 1.0   \\
        \hline
        \hline
    \end{tabular}
	
\end{table}

\begin{figure*}
	\centering
	\includegraphics[width = 16 cm]{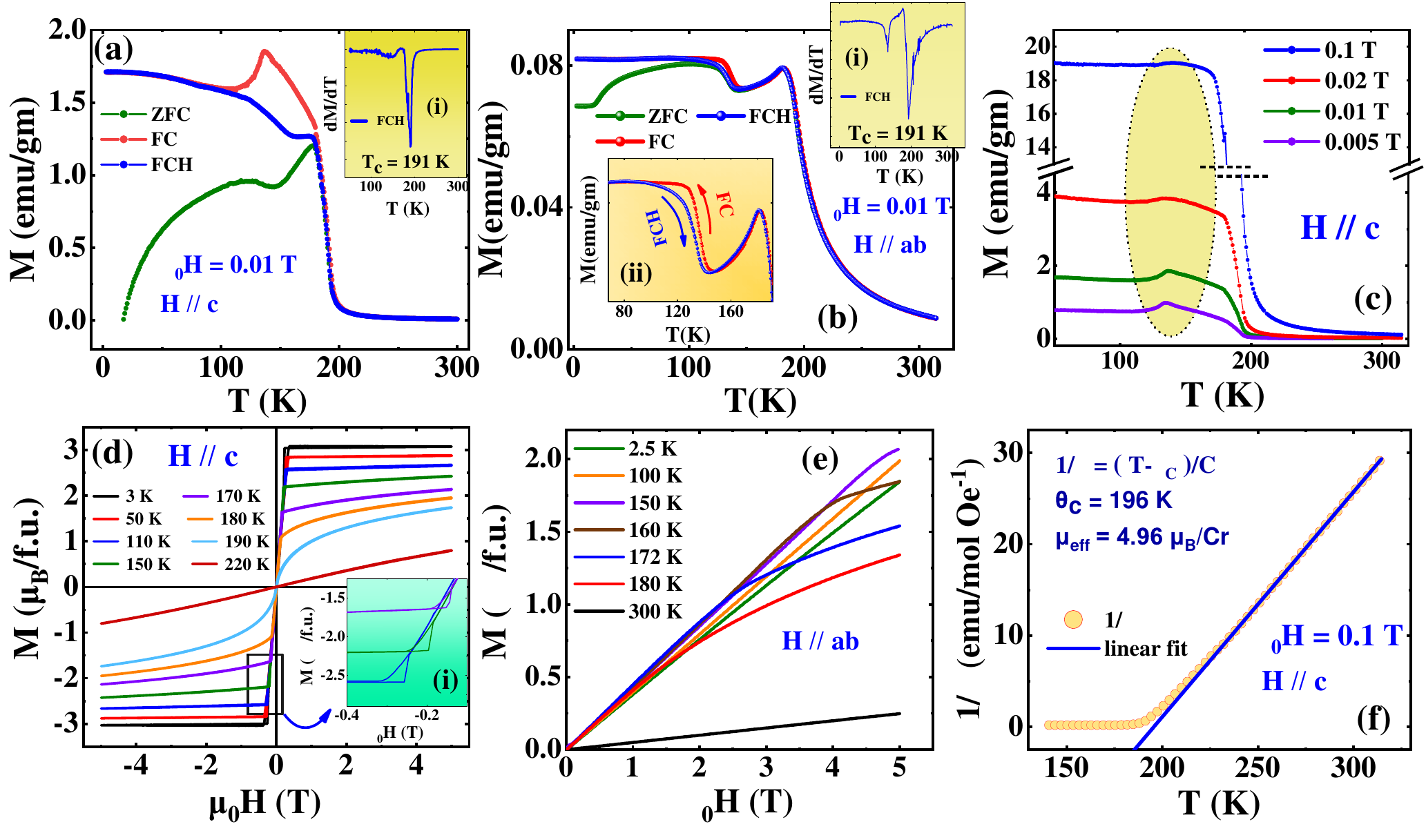}
	\caption{ Temperature ($T$) dependence of magnetization ($M$) in zero field-cooled-heating (ZFC), field-cooled (FC) and field-cooled-heating (FCH) protocols under  $\mu_0H$= 0.01 T and the inset (i) shows the temperature derivative of $M(T)$ for (a) $\mu_0H\parallel c$ and (b) $\mu_0H\perp c$. (c) Temperature ($T$) dependence of magnetization ($M$) under different fields ($\mu_0H$=0.005 T, 0.01 T, 0.02 T, 0.1 T) for $\mu_0H\parallel c$. (d) and (e) Represents isothermal magnetization at different temperatures measured between $\pm$5 T and 0-5 T for $\mu_0H\parallel c$ and $\mu_0H\perp c$ respectively. (f) Inverse susceptibility ($\chi$$^{-1}$) versus temperature ($T$) data with linear fitting, measured at $\mu_0H$ = 0.1 T for $H\parallel c$.} 
	\label{fig:MT}
\end{figure*}

\subsection{Magnetization}

The temperature ($T$) dependence of magnetization ($M$) of Cr$_{1+\delta}$Te$_2$, $\delta \sim$ 0.33, measured in zero field-cooled-heating (ZFC), field-cooled (FC) and field-cooled-heating (FCH) protocols, is shown in Fig.~\ref{fig:MT} (a) and Fig.~\ref{fig:MT} (b) with magnetic field ($\mu_0H$ = 0.01 T) applied along the  $c$-axis ($\mu_0H\parallel c$) and the $ab$ plane ($\mu_0H\perp c$), respectively. The insets of Fig.~\ref{fig:MT} (a) and (b) display the $dM/dT$ vs $T$ plots for $\mu_0H\parallel c$ and $\mu_0H \perp c$, respectively. From the minima of the $dM/dT$ vs $T$ curves, the paramagnetic to ferromagnetic (FM) transition temperature ($T_C$) is found to be around 191 K for both orientations. The large difference in magnetization measured along the $c$-axis and the $ab$-plane indicates the presence of strong magnetic anisotropy. For $\mu_0H\parallel c$, $M(T)$  shows large thermomagnetic irreversibility between ZFC and FC branch below $T_C$, and it is more prominent below 100 K, which can be attributed to  magneto-crystalline anisotropy~\cite{Anisotropy} and/or spin canting~\cite{AHE_mono_Cr5Te8, Cr0.62Te}. On the other hand,  relatively weaker thermomagnetic irreversibility occurs below 100 K between ZFC and FC branch for the \emph{ab}-plane.
\par
When measured along $\mu_0H\parallel c$ [Fig.~\ref{fig:MT} (a)], FC and FCH curves show large thermal hysteresis between 180-110 K with a peak in the FC curve at 140 K. This peak broadens with increasing value of $\mu_0H$ [Fig.~\ref{fig:MT} (c)].  The thermal hysteresis, albeit much weaker, is also present when measured along $\mu_0H\perp c$ [Fig.~\ref{fig:MT} (b)]. The presence of thermal hysteresis indicates first order phase transition between two different magnetic phases or due to some magneto-structural transition. Similar hysteresis was observed in few other layered compounds, such as Cr$_5$Te$_8$~\cite{AHE_mono_Cr5Te8}, Fe$_3$GeTe$_2$~\cite{Fe3GeTe_Hysteresis,Fe3GeTe2_PRB}.

\par
Isothermal $M$ vs $\mu_0H$ data recorded at different constant $T$, between $\pm$5 T, for both $\mu_0H\parallel c$ and $\mu_0H\perp c$ are shown in Fig.~\ref{fig:MT} (d) and Fig.~\ref{fig:MT} (e), respectively. $M$ vs $\mu_0H$ data for $\mu_0H\parallel c$  show clear FM behaviour below  $T_C$, and $M$ saturates at an applied field of 0.35 T (for $\mu_0H\parallel c$), while $M$ shows nonsaturating behaviour for $\mu_0H\perp c$, even at an applied field of 5 T. This indicates the existence of strong magnetic anisotropy with the  $c$ axis being the easy magnetization direction. For $\mu_0H\parallel c$, saturation magnetization ($M_s$) is found to be 3.08 $\mu_B/f.u.$ ($\sim$2.33 $\mu_B$/Cr) at 3 K, which is less compared to the free Cr$^{3+}$ ion ($\sim$3 $\mu_B$/Cr). This deviation may be due to the itinerant nature of the Cr-$3d$ electrons, mixed valent nature of Cr or spin canting~\cite{Neutron_Cr5Te8, AHE_mono_Cr5Te8,Canted1, Canted2,Jpcm_structure,Cr5Te8_MCE}. Interestingly, we observe a staircase like nature of the magnetization curve when $M$ drops from its saturation value towards zero with the decrease of $\mu_0H$. This staircase effect is only visible when measured for $\mu_0H\parallel c$, and it is shown in the inset of Fig.~\ref{fig:MT} (d). At low $T$ ($\sim$ 3-120 K), this staircase like jump occurs around $\mu_0H_{jump} \sim$ 0.28-0.24 T and $\mu_0H_{jump}$ decreases to lower field with increasing $T$. Similar behaviour has also been observed in some Cr-Te based compounds~\cite{AHE_mono_Cr5Te8,Adv.Mat}, where it was assigned to spin-flop transition from a non-collinear to a collinear FM state. Interestingly, Saha \textit{et al.}~\cite{NatCom_RanaSaha} reported a similar staircase effect in the Hall resistivity versus $\mu_0H$ curve of Cr$_{1.33}$Te$_2$, although it is absent in their magnetization data. Staircase effect in the magnetization data is not uncommon in FM-like materials such as PrMn$_2$Ge$_2$~\cite{PrMnGe}, BaFe$_{12}$O$_{19}$~\cite{BaFeO}, which is typically caused by the movement of domain walls. 

\par
We have examined the  $\chi$$^{-1}$ vs $T$ data measured at $\mu_0H$= 0.1 T for $\mu_0H\parallel c$ in the high-$T$ range, and it is shown in Fig.~\ref{fig:MT} (f). $\chi$$^{-1}$ is fitted in the $T$ range 240-315 K with Curie-Weiss (CW) law $\chi$$^{-1}=(T-\theta_C)/C$, and the obtained CW temperature is $\theta_C=$196 K and the effective magnetic moment is found to be $\mu_{eff} \sim 4.96~\mu_B$/Cr. Large positive value of $\theta_C$ indicates, ferromagnetic exchange interaction is dominating in Cr$_{1+\delta}$Te$_2$.

The anisotropic magnetic properties in Cr$_{1+\delta}$Te$_2$ is further explored through magnetic entropy change $\Delta S_M(T,H)$, which is a measure of magneto-caloric effect of the sample~\cite{FRANCO2018112}. It can be obtained using a numerical approach based on Maxwell relation to the equation $\Delta S_M(T,H) = \int_{0}^{H}(\frac{dM}{dT})_HdH$. For magnetization measured in small discrete field and $T$ intervals, the above formula can be approximated as 

\begin{equation}
\Delta S_M(T_i,H) = \frac{\int_{0}^{H}M(T_i,H)dH-\int_{0}^{H}M(T_{i+1},H)dH}{T_i-T_{i+1}}
\label{eqn:DeltaS}
\end{equation}

\begin{figure}[ht]
	\centering
	\includegraphics[width = 8 cm]{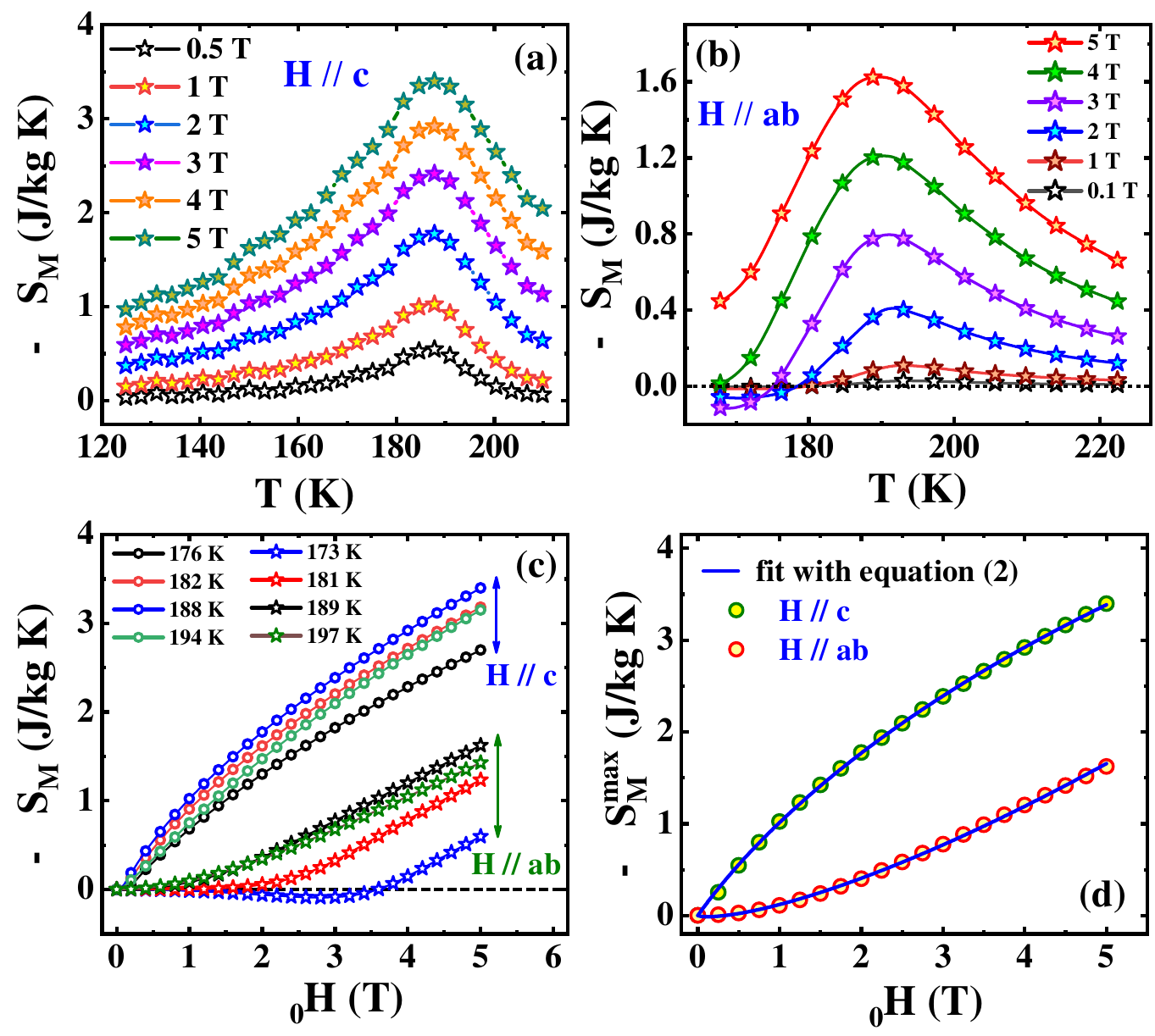}
	\caption{ (a) Temperature dependence of magnetic entropy change ($-\Delta S_M$) (a) along $c$ axis and (b) in the $ab$ plane. (c) Field dependence of entropy changes $-\Delta S_M$ at different $T$ along $c$ axis and $ab$ plane. (d) Field dependence of maximum entropy change ($-\Delta S_M^{max}$) along $c$ axis and $ab$ plane.}
	\label{fig:MCE}
\end{figure}

The calculated $-\Delta S_M(T,H)$ as a function of $T$ in field up to 5 T for $\mu_0H\parallel c$ and $\mu_0H\perp c$ are shown in Fig.~\ref{fig:MCE} (a) and Fig.~\ref{fig:MCE} (b) respectively. The obtained $-\Delta S_M$ shows maxima in the vicinity of transition temperature with a maximum value of 3.39 J Kg$^{-1}$ K$^{-1}$ for $\mu_0H\parallel c$ and 1.62 J Kg$^{-1}$ K$^{-1}$ for $\mu_0H\perp c$ at 5 T of applied field. Interestingly, $-\Delta S_M$ is found to be positive over the full $T$ range (125-210 K) for $\mu_0H\parallel c$, while it turns negative for $\mu_0H\perp c$ at low $T$ ($<$ 175 K) for field below 3.5 T. This change in sign is visible in the $-\Delta S_M$ vs $\mu_0H$ plot with $\mu_0H \perp c$, however it is absent when measured along the $c$ axis [Fig.~\ref{fig:MCE} (c)]. The field dependence of maximum entropy $-\Delta S_M^{max}$ is shown in Fig.~\ref{fig:MCE} (d) for both $\mu_0H\parallel c$ and $\mu_0H\perp c$. Evidently, $-\Delta S_M$ is highly anisotropic, which is in line with the anisotropic magnetic nature of the compound.  
\par
For an FM system undergoing a second-order phase transition, $-\Delta S_M^{max}$ was predicted to grow as $H^{2/3}$~\cite{Smax1}, but it holds good only for a high magnetic field. A more practical $H$-dependence of $-\Delta S_M^{max}$ is obtained on the basis of Landau’s theory of second-order phase transitions~\cite{Landau1,Landau2}:
\begin{equation}
-\Delta S_M^{max} = A(H+H_0)^{2/3}-AH_0^{2/3}+BH^{4/3}
\label{eqn:DeltaSH}
\end{equation}
Here, $A$ and $B$ are intrinsic parameters and $H_0$ is a parameter dependent on purity and homogeneity. Our fitting to the data [see Fig.~\ref{fig:MCE} (d)] yields
$A$=1.79(9) JKg$^{-1}$K$^{-1}$T$^{-2/3}$, $B=-$0.10(3) JKg$^{-1}$K$^{-1}$T$^{-4/3}$ for $H\parallel c$. Similarly, for $H\perp c$, $A=-$0.11(5) JKg$^{-1}$K$^{-1}$T$^{-2/3}$, $B$=0.23(3) JKg$^{-1}$K$^{-1}$T$^{-4/3}$. Notably, both $A$ and $B$ have opposite signs for $H\parallel$ and $\perp c$.

\subsection{Structural Study}

\begin{figure}
	\centering
	\includegraphics[width = 7.5 cm]{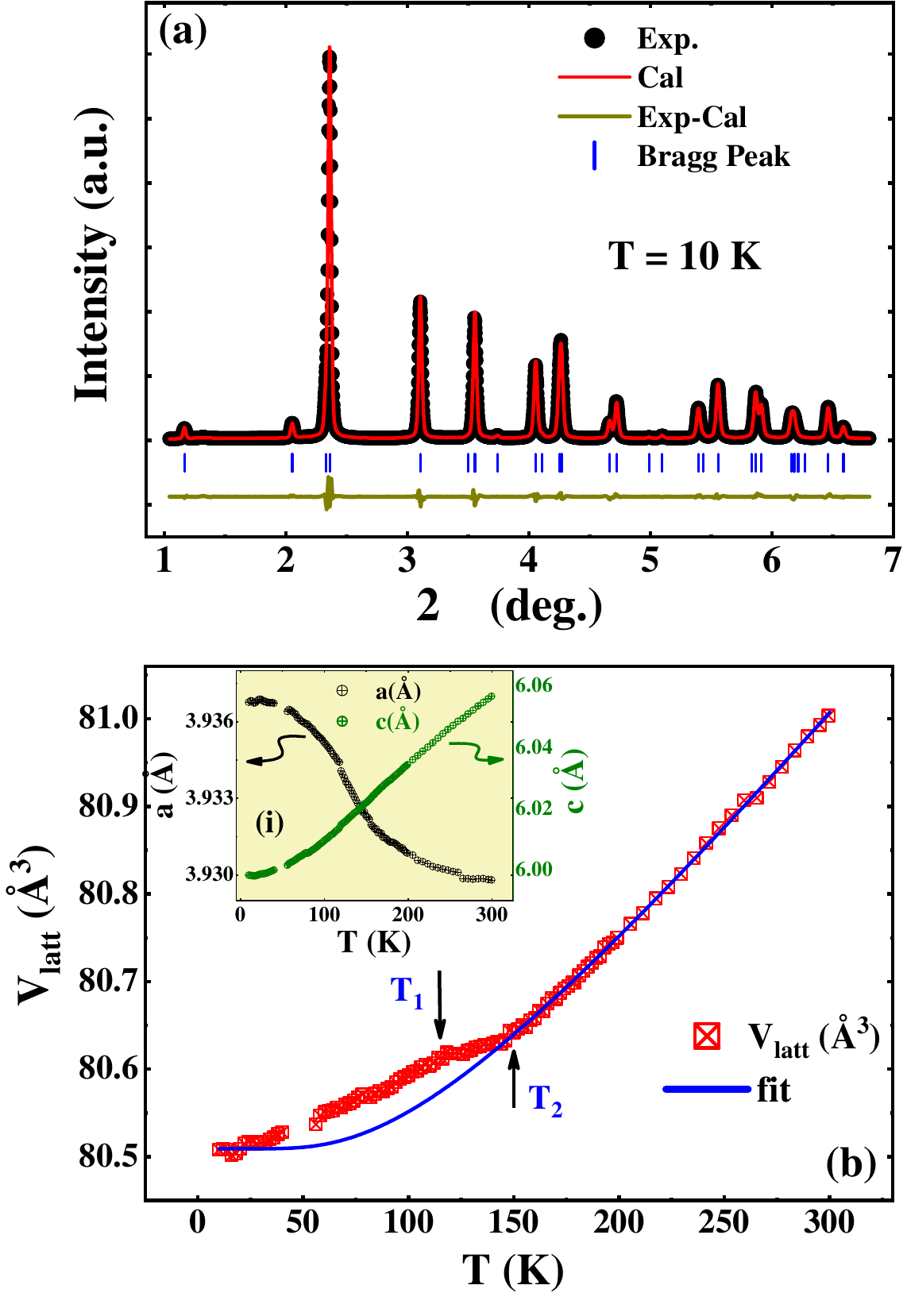}
	\caption{ (a) PXRD data along with Rietveld refinement at 10 K. (b) Shows the $T$ variation of unit cell volume along with the fitting with eqn.~\ref{eqn:Volume}, inset (i) shows the $T$ dependence of lattice parameter $a$ and $c$.}
	\label{fig:lattice}
\end{figure}

The significant thermal hysteresis observed in the $M$ vs $T$ data prompts us to explore the $T$-dependent evolution of the crystal structure. The structural properties are investigated through synchrotron-based PXRD in the $T$ range 10-300 K. Rietveld analysis shows that the sample retained its room temperature symmetry down to at least 10 K [see Fig.~\ref{fig:lattice} (a)]. The $T$ variation of the lattice parameters $a$ and $c$ are shown in the inset of Fig.~\ref{fig:lattice} (b). $a$ shows negative thermal expansion (NTE) behaviour throughout the $T$ range with a relatively rapid change between 150-100 K, while lattice parameter $c$ shows positive thermal expansion (PTE) down to 10 K. Interestingly, unit cell volume ($V_{latt}$) shows nearly zero thermal expansion (NZTE) behavior between 115 K ($=T_1$) and 150 K ($=T_2$), and PTE below $T_1$ as well as above $T_2$ [see Fig.~\ref{fig:lattice} (b)]. The lattice parameter $a$ decreases sharply during heating from $T_1$ to $T_2$, while the lattice parameter $c$ increases in the same process, which eventually leads to almost no change in $V_{latt}$ between $T_1$ and $T_2$.

In general, the thermal variation of $V_{latt}$ can be expressed as~\cite{Surajit}

\par
\begin{equation}
V_{latt}(T) = V_0[1+\frac{\kappa}{(e^{\theta_D/T}-1)}]	
	\label{eqn:Volume}
\end{equation}
\noindent
where $V_0$ is the unit-cell volume at 0 K, $\theta_D$ is the Debye temperature and $\kappa$ is the fitting parameter. As depicted in Fig.~\ref{fig:lattice} (b), $V_{latt}$ obeys eqn.~\ref{eqn:Volume} quite well down to 150 K, and below which it started to deviate from the experimental data. The obtained fitting parameters are $V_0$ = 80.5(1) \AA$^3$, $\kappa$=0.01(1) and $\theta_D$= 310 K.

The observed temperature region of NZTE falls within the $T$ range where thermal hysteresis has been observed in the magnetization data. This points towards the existence of magneto-structural coupling in the system. Recently~$\emph{Chen Li et.al.}$ reported diverse thermal expansion behaviour in ferromagnetic Cr$_{1-\delta}$Te and they have shown that with increasing  Cr-concentration, $V_{latt}$ changes from PTE to NZTE, and eventually to NTE~\cite{ZTE}.  

\subsection{Electrical Resistivity \& Magnetoresistance}

\begin{figure}
	\centering
	\includegraphics[width=8cm]{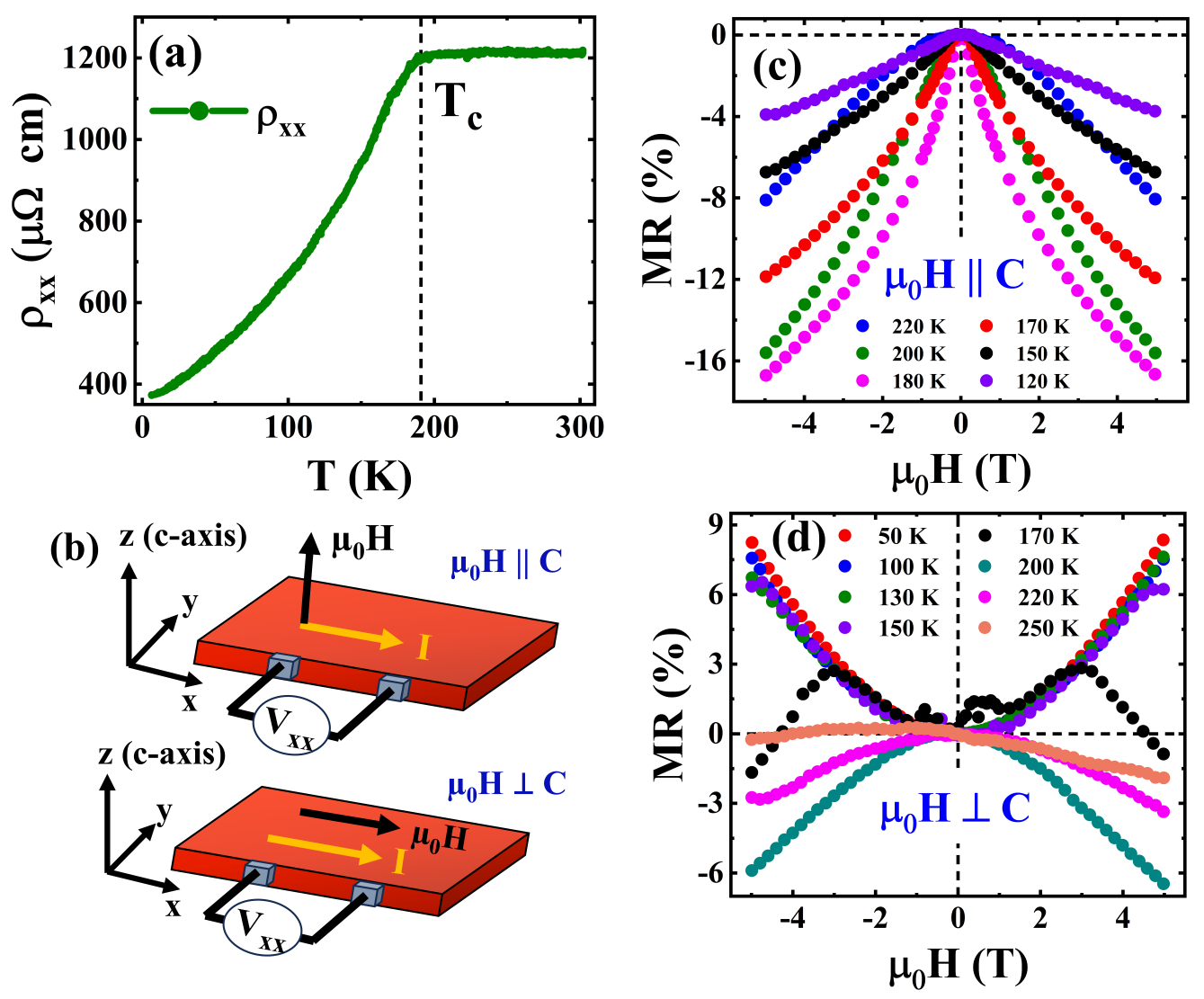}
	\caption{ (a) $T$ dependence of longitudinal resistivity ($\rho _{xx}$). (b) Schematic diagram from MR measurements. (c) Shows MR at different constant $T$ for $\mu_0H\parallel c$. (d) MR vs $\mu_0H$ at different $T$ for $\mu_0H\perp c$.} 
	\label{fig:RT}
\end{figure}

Fig.~\ref{fig:RT} (a) shows the $T$ dependence of in-plane $\rho _{xx}$ measured between 6-300 K. The sample shows metallic behaviour down to the lowest temperature with a residual resistivity ratio (RRR) = $\rho (300 K)/\rho (6 K)$ = 3.26. A clear slope change is observed around 191 K, which indicates the $T_C$ in the system. Below $T_C$, $\rho _{xx}$ decreases rapidly with $T$ due to the loss  in spin-disordered scattering.
\par
Next, we measured the isothermal $\mu_0H$ dependence of $\rho _{xx}$ for different constant $T$. Fig.~\ref{fig:RT} (c) and (d) depicts the magnetoresistence (MR), defined as MR $= [\rho_{xx}(\mu_0 H) - \rho_{xx}(0))]/\rho_{xx}(0) \times 100\%$, measured for field applied along and perpendicular to the $c$-axis respectively. Fig.~\ref{fig:RT} (b) shows the schematic of the measurement configuration for both directions. For $\mu_0H\parallel c$, unsaturated negative MR has been observed for all the $T$, with a maximum value of $\sim$ 17 \% near $T_C$ [Fig.~\ref{fig:RT} (c)]. In  case of FM intermetallic compounds containing 3$d$ transition metal, the magnetic contribution to the resistivity arises due to the scattering of the delocalized $s$ electrons with the partially localized 3$d$ electrons. As the $c$-axis is the easy axis of magnetization, the spin disorder scattering is greatly suppressed under an applied $\mu_0H \parallel c$, leading to a negative MR~\cite{sd1,sd2}. The spin disorder scattering is prominent near $T_C$ and decreases with a decrease in $T$ and it is reflected in the MR data. Fig.~\ref{fig:RT} (d) shows MR for $\mu_0H\perp c$ configuration with $H$ applied parallel to the current ($I$). Above $T_C$, the sample exhibits negative MR, but its magnitude is much lower than that observed for $H$ parallel to the $c$-axis, indicating strong anisotropy in the magneto-transport~\cite{Aniso_MR}. Below $T_C$, positive MR has been observed with a non-saturating maximum value of $\sim$ 8\% at $\pm$5 T as the $ab$-plane is the hard axis of magnetization.

\subsection{Hall Resistivity}

\begin{figure*}[ht]
	\centering
	\includegraphics[width = 16 cm]{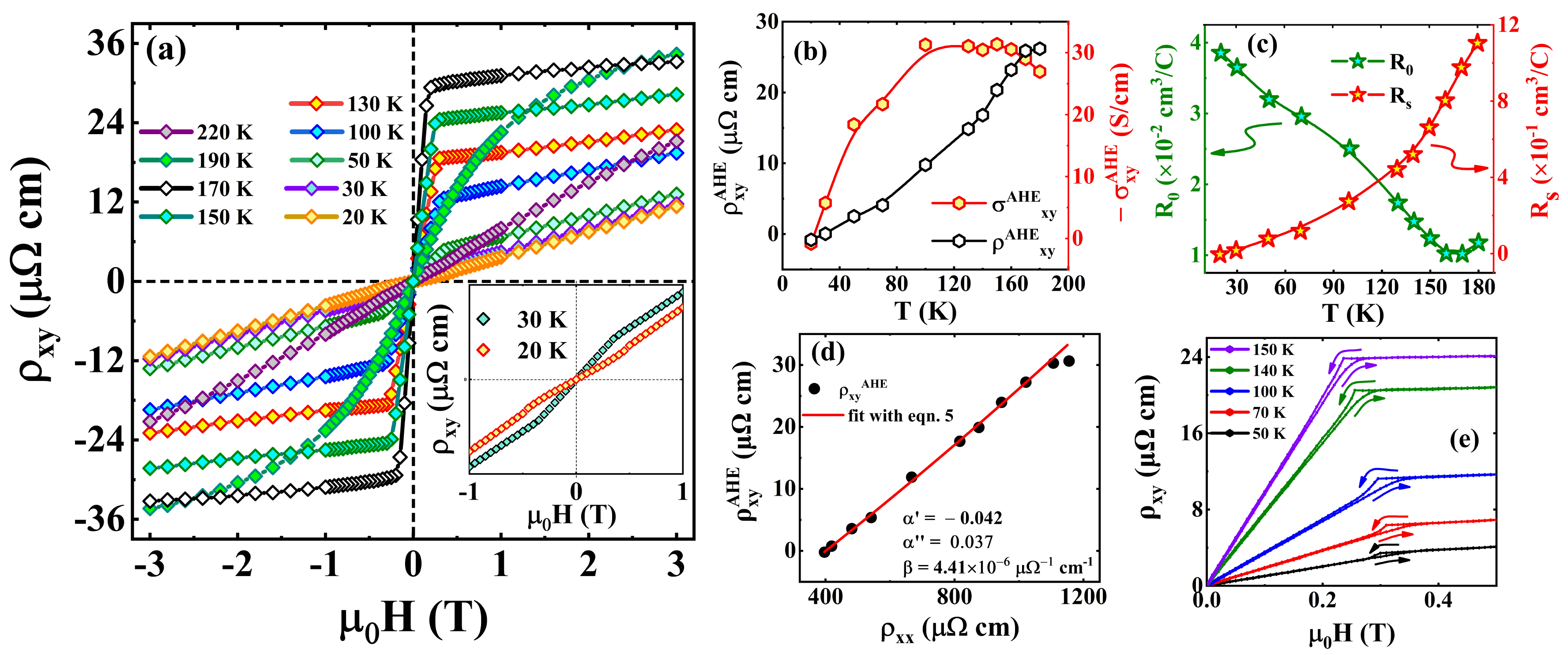}
	\caption{ (a) Hall resistivity $\rho_{xy}$ vs $\mu_0H$ at different constant $T$ for $\mu_0H\parallel c$ and the inset shows the enlarge view of the low $T$ data. (b) $T$ variation of $\rho_{xy}^{AHE}$ and corresponding $\sigma_{xy}^{AHE}$. (c) The left and right panel shows $T$ variation of ordinary Hall coefficient ($R_0$) and anomalous Hall coefficient ($R_s$) respectively. (d) The plot of $\rho_{xy}^{AHE}$ vs $\rho_{xx}$ with a fitting using eqn.~\ref{eqn:Scaling}. (e) $\rho_{xy}$ vs $\mu_0H$ plot in the low field region. }
	\label{fig:Hall}
\end{figure*}

The Hall resistivity ($\rho _{xy}$) for Cr$_{1+\delta}$Te$_2$ measured as a function of $\mu_0H$ ($\mu_0H$ applied along the $c$ axis) at different constant $T$ [Fig.~\ref{fig:Hall} (a)]. To eliminate the longitudinal resistivity contribution due to the misalignment of voltage leads, $\rho _{xy}$ was calculated using the formula $\rho_{xy}(\mu_0H) = \frac{1}{2}[\rho_{xy}(+\mu_0H) - \rho_{xy}(-\mu_0H)]$. The $\rho _{xy}$ shows highly nonlinear behaviour with $\mu_0H$ below $T_C$: it increases sharply in the low field region and shows a saturating tendency at higher fields, indicating the presence of AHE in the system. 
\par
Typically in FM materials, $\rho _{xy}$ can be expressed by an empirical formula~\cite{Hall1_PRB,AHE_mono_Cr3Te4},

\begin{equation}
\rho_{xy} = \rho_{xy}^{OHE} + \rho_{xy}^{AHE} = R_0\mu_0H + R_s \mu_0M	
	\label{eqn:Hall}
\end{equation}

where, $\rho_{xy}^{OHE}$ is the ordinary Hall resistivity related to Lorentz effect and $\rho_{xy}^{AHE}$ is the anomalous Hall resistivity. $R_0$ and $R_s$ represent the ordinary and anomalous Hall coefficient respectively, which can be obtained by linear extrapolation of the high-field $\rho _{xy}(\mu_0H)$ data to $\mu_0H=$ 0 T~\cite{Hall1_PRB}. 
\par
Fig.~\ref{fig:Hall} (b) shows the $T$ dependence of the obtained $\rho_{xy}^{AHE}$ and it turns negative at low-$T$. The corresponding anomalous Hall conductivity (AHC), $\sigma_{xy}^{AHE}$ ($= -\rho_{xy}^{AHE}/[(\rho_{xy}^{AHE})^2+(\rho_{xx})^2]$) is shown in the right axis of Fig.~\ref{fig:Hall} (b). One can see that the magnitudes of both $\rho_{xy}^{AHE}$ and $\sigma_{xy}^{AHE}$ increase with increasing $T$, and $\sigma_{xy}^{AHE}$ shows a tendency towards saturation as $T$ approaches $T_C$. The left and right axes of Fig.~\ref{fig:Hall} (c) show $T$ variation of $R_0$ and $R_s$ respectively. Notably, $R_0$ is positive at all $T$, indicating holes are the majority charge carriers, and the carrier concentration is found to be 1.6-5.3$\times$10$^{20}$ cm$^{-3}$ in the $T$ range 20-180 K. The magnitude of $R_s$  is one order of magnitude larger than $R_0$, indicating the dominance of anomalous Hall resistivity.

\par
Three primary mechanisms are responsible for the AHE, and they follow a scaling relation $\rho_{xy}^{AHE} \propto \rho_{xx}^q$, with $q$ = 1 for skew-scattering and $q$ = 2 for intrinsic and side-jump contributions. Recently a modified scaling law has been proposed~\cite{PRL_Scaling1,PRL_Scaling2}:

\begin{equation}
\rho_{xy}^{AHE} = \alpha^{\prime} \rho_{xx0} + \alpha^{\prime \prime} \rho_{xxT} + \beta \rho_{xx}^2 	
	\label{eqn:Scaling}
\end{equation}

\noindent Here $\alpha^{\prime} \rho_{xx0}$ is the residual resistivity contribution which is caused by static impurities and $\alpha^{\prime \prime}\rho_{xxT}$ is the phonon induced scattering contribution. $\beta$ represents the coefficient for side-jump and/or intrinsic AHE. To separate the different contributions to AHE, we have plotted $\rho_{xy}^{AHE}$ as a function of $\rho_{xx}$ [see Fig.~\ref{fig:Hall} (d)] and fitted it with eqn.~\ref{eqn:Scaling}. From the fitting we obtained, $\alpha^{\prime}$ = - 0.042, $\alpha^{\prime \prime}$ = 0.037, $\beta$ = 4.41$\times$10$^{-6}$ $\mu\ohm^{-1}$cm$^{-1}$ = 4.41 S cm$^{-1}$. The analysis indicates that the AHE is primarily governed by the skew scattering mechanism, with negligible contributions from the side-jump effect and Berry curvature-induced intrinsic mechanisms.
\par
Similar to the $M$ vs $\mu_0H$ data for $\mu_0H \parallel c$, the staircase like jump has also been observed in our $\rho _{xy}$ vs $\mu_0H$ plots in the similar field region [see Fig.~\ref{fig:Hall} (e)]. This further demonstrates the correlation between $M$ and AHE, which may be attributed to a magnetization-induced spin-flop process~\cite{AHE_mono_Cr5Te8}. Previously Saha \textit{et al.}~\cite{NatCom_RanaSaha} assigned this jump in $\rho _{xy}$ to THE, as such jumps were absent in their magnetization data. The absence of THE  possibly due to the centro-symmetric crystal structure of our sample.


\section{Ab initio calculation}

Recent study of Cr$_{1+\delta}$Te$_2$ materials is speculated to have intrinsic AHE due to finite BC contribution~\cite{Fujisawa}, whereas, our experimental data suggest the absence of an intrinsic AHE. To address this discrepancy, we performed electronic band structure calculations for the compound. We constructed a 
supercell to match the experimental composition of the compound, finding that Cr$_{10}$Te$_{16}$ (= Cr$_{1.25}$Te$_2$) 
provides a close approximation to the experimental structure. It was suggested by  P. Li \textit{et al.}~\cite{P.Li} that the
origin of the intrinsic AHE and magnetism are strongly correlated, therefore, we explore the origin of the 
magnetic properties of Cr$_{1.25}$Te$_2$ theoretically.
\par
\begin{figure*}
\centering
\includegraphics[width=0.75\linewidth]{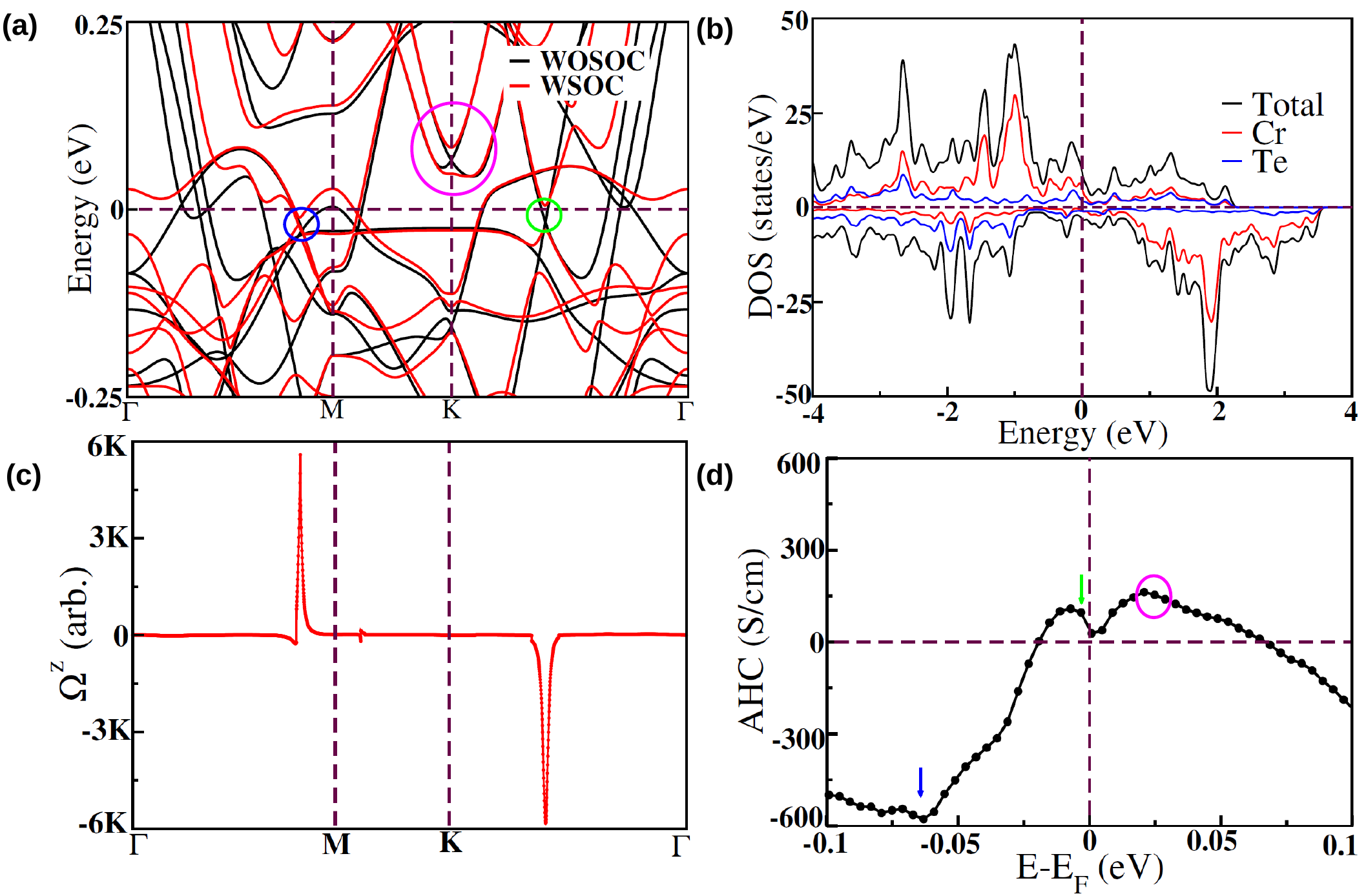}
\caption{(a) The band structure of Cr$_{1.25}$Te$_2$ without SOC and with SOC in red and blue color respectively. The blue, and green circles denote the nodes below $E_F$, while the magenta one denotes a node in the unoccupied conduction band. (b) Density of states of Cr$_{1.25}$Te$_2$. (c) The Berry curvature along the high symmetry lines due to the nontrivial crossings. (d) Energy (\textit{E-E$_F$}) dependence of the AHC for Cr$_{1.25}$Te$_2$.}
\label{fig:band}
\end{figure*}

Fig.~\ref{fig:band} (b) illustrates the computed density of states (DOS), where the black solid line represents the total DOS of the compound. The projected density of states for the $d$-orbitals of Cr and $p$-orbitals of Te are shown as red and blue lines respectively in Fig.~\ref{fig:band} (b). 
The spin-up and spin-down $d$-orbitals of Cr look quite asymmetric, with a sharp peak in the spin-up orbitals around -1 eV below the fermi-energy ($E_F$) in DOS, indicating localized spin-up moments. In contrast, the finite DOS of the spin-down band around -2 eV suggests a small localized moment. 
The finite DOS of the spin-up orbital and the much lower DOS of the spin-down orbital at $E_F$ indicate that spin-up electrons contribute mostly to magnetism. The magnetic moment of the system is largely due to the $d$-electrons of the Cr atoms and Te exhibits 
negligible magnetic moments as their up and down spin orbital have almost the same contribution. The overall magnetic moment of Cr$_{1.25}$Te$_2$ is calculated to be 3.08 $\mu_B$ per formula unit, which is very similar to the experimentally observed value. 
\par         

To understand the topological characteristics of the electronic states and their impact on transport behavior, we have employed a plane wave-based pseudopotential method to compute the band dispersion, both with and without spin-orbit coupling (SOC). The analysis begins by examining the electronic structure in the absence of SOC. In this scenario, nontrivial nodal points representing band crossings are observed along the two high-symmetry paths: $\Gamma$-M and $\Gamma$-K, both near E$_F$. These nodal points, where the black solid lines cross, are highlighted by green and blue circles in Fig.~\ref{fig:band} (a), indicating their significance in the band structure. When SOC is incorporated into the calculations, together with the appropriate magnetic ordering aligned along the compound's easy axis, the previously identified band-crossing points disappear, and a gap opens up at those points. In Fig.~\ref{fig:band} (a) the modified band dispersions in the presence of SOC are represented by a red solid line, clearly showing the emergence of gapped nodal points near $E_F$.

\par
To further explore the consequences of these gapped nodal points on electron dynamics, we have calculated the BC of the Bloch wave function in the momentum space~\cite{Jungwirth}. BC is a crucial quantity in topological materials, functioning as a pseudo-magnetic field that can influence electron motion, particularly by generating an anomalous transverse velocity. Next, we extracted Wannier functions from the DFT band structure with the WANNIER90 package~\cite{pizzi2020wannier90,MarzariPhysRevB.56.12847}, using the $s$ and $d$ orbitals of Cr and the $s$ and $p$ orbitals of Te for initial projections. From these projections, we have constructed a tight-binding Hamiltonian $\mathcal{H}$  using the Wannier functions and calculated the component of BC along the $z$ direction ($\Omega^{z}_n$ ) using the relation~\cite{xiao2010berry}:

 \begin{eqnarray}
\Omega^z_{n} = -2i \sum_{m \neq n} \frac{{\langle \psi_{n\textit{k}}|v_x|\psi_{m\textit{k}}\rangle} {\langle \psi_{m\textit{k}}|v_y|\psi_{n\textit{k}}\rangle}} {[E_{m}(\textit{k}) - E_{n}(\textit{k})]^2}
\end{eqnarray} 
Here $n$ is the occupied band's index, $E_{n}(\textit{k})$ is the eigenvalue of the $n$th eigenstate $\psi_{n}(\textit{k})$, $v_i$ = $\frac{1}{\hbar}\frac{\partial H(\textit{k})}{\partial \textit{k}_i }$ is the velocity operator along the $i$ ($i =x,y,z$) direction.
\par
In Fig.~\ref{fig:band} (c), the calculated BC is shown for the gapped nodal points. Specifically, a positive BC is observed along the high-symmetry direction $\Gamma$-M, while a negative BC is present along the high-symmetry direction $\Gamma$-K. These findings emphasize the impact of SOC on the compound’s electronic structure and its resulting topological properties. To gain a clearer understanding of the two distinct crossing points mentioned earlier, we have performed fat band calculations to analyze the specific orbital contributions to the bands near these points. The crossing point marked by the blue circle in Fig.~\ref{fig:band} (a), located along the $\Gamma$-M path, arises due to the hybridization of two $d$-orbitals of Cr. As we move from $\Gamma$ to M, the symmetry of the system is reduced from a higher symmetry to C$_{2v}$, influencing the nature of the band structure and the crossing point.
\par
In contrast, the green circle surrounding the other crossing point, which occurs along the $\Gamma$-K path, results from the interaction between the $d$-orbital of Cr and the $p$-orbital of Te. As we move from $\Gamma$ to K, the symmetry decreases from D$_{6h}$ to D$_{3h}$. This symmetry reduction and the mixing of different orbital types (Cr $d$-orbital and Te $p$-orbital) play a pivotal role in shaping the band structure at this crossing point.

\par
The symmetry properties and the spatial distribution of the wave functions can significantly influence the BC, especially in cases where the crossing points involve 3$d$-3$d$ (Cr-Cr) and 3$d$-5$p$ (Cr-Te) mixing. The interplay between these orbitals and their respective symmetries can lead to a distinct pattern in BC, which may even shift its sign. BC is directly proportional to the AHC, and a significant BC can result in a notable AHC. To quantify the intrinsic AHC ($\sigma_{xy}^{int}$), we can employ the Kubo formalism's linear response theory~\cite{Gradhand_2012}. For the \textit{xy} plane, the expression for the AHC is as follows: 

\begin{equation}
 \sigma_{xy}^{int} = -{\frac{e^2}{\hbar} \int\frac{d^{3}\textit{k}}{(2\pi)^3}\sum_{n}\Omega^z_{n}(\textit{k})f_{n}(\textit{k})}  
 \label{eqn:kubo}
 \end{equation}

where $f_{n}(\textit{k})$ represents the Fermi-Dirac distribution function. Despite the substantial BC values, their opposing signs lead to near-cancellation when calculating the AHC which is shown in Fig.~\ref{fig:band} (d). It highlights two distinct energy positions, marked by blue and green arrows, where the AHC sign flips. This behavior is a direct consequence of the BC sign reversal, which arises from changes in the band topology at these specific energy levels. It, therefore, becomes evident that  \textit{even though the compound exhibits nontrivial band topology around $E_F$, its intrinsic AHC remains effectively zero}.
\par
Beyond the two nodal points in the valence band, an additional nodal point emerges in the conduction band at the K point of the Brillouin zone, marked by a magenta circle as shown in Fig.~\ref{fig:band} (a). To unravel its origin, we delve into the crystalline symmetry of the Cr-Te bilayer, which is defined by a three-fold rotation around the $z$-axis (C$_{3z}$), a two-fold rotation around the $y$-axis (C$_{2y}$), and inversion symmetry ($\cal{P}$). Remarkably, the system remains invariant under the symmetries C$_{3z}$, C$_{2y}$, and the combined $\cal{PT}$ symmetry, where $\cal{T}$ represents time-reversal symmetry. These symmetry properties give rise to two-dimensional irreducible representations, culminating in a two-fold degeneracy at the K point in the absence of SOC~\cite{Kim2018} as illustrated by the magenta circle in Fig.~\ref{fig:band} (a). However, the introduction of SOC disrupts this degeneracy, resulting in a pronounced anomalous electron velocity [see Fig.~\ref{fig:band} (a)]. If the electronic states in this region near the K point are occupied, the system could manifest a significant intrinsic anomalous Hall effect [see Fig.~\ref{fig:band} (d)].

\section{Summary and Conclusion}

The work presents a comprehensive experimental and theoretical investigation on the self intercalated Cr$_{1+\delta}$Te$_2$ compound with $\delta \sim$ 0.33. The compound has a layered trigonal crystal structure, with the stacking of layers along the crystallographic $c$-axis. Our magnetic and magneto-transport measurements show strong anisotropy both in $M$ and $\rho_{xx}$ with the easy axis of magnetization along the $c$-direction. The composition also shows strong anisotropy in the magneto-caloric effect with the magnitude of $\delta S$ being higher for $\mu_0H \parallel c$. 

\par
An important outcome of the study is the observation of thermal hysteresis between heating and cooling curves of $M$ versus $T$ data. Such hysteresis is an indication of first order magnetic transition. We also observe an iso-structural lattice anomaly in the region of thermal hysteresis, which indicates the coupling between magnetic moment and lattice.

\par
The compound, akin to other members of the Cr-Te family, shows a significant value of AHE. Our analysis of the Hall data indicates that the major contribution comes from the extrinsic skew-scattering phenomenon. The complete absence of an intrinsic BC mechanism is rather astounding considering the manifestation of topological aspects both in real and momentum spaces in similar layered metal chalcogenides. 

\par
We perform the \emph{ab} initio electronic structure calculation to understand the absence of intrinsic AHE in the system. We note the presence of nodal points (band crossing) close to $E_F$ along high symmetry directions in the momentum space. In the presence of magnetic order along the axial direction and the presence of SOC, the degeneracies at the nodal point are lifted and the DFT + SOC calculation clearly indicate the presence of non-zero BC values, although having both positive and negative signs at different crossing points. The symmetry properties and spatial distribution of wave functions play a crucial role in determining the phase in the BC, particularly in cases of 3$d$-3$d$ (Cr-Cr) and 3$d$-5$p$ (Cr-Te) orbital mixing. This interaction can lead to distinct patterns in the BC, with the potential for a sign shift depending on the nature of the orbital symmetries involved. Despite these significant BC values, they nearly cancel each other out, resulting in a negligible intrinsic AHE for the system. Notably, Fujisawa \textit{et al.}~\cite{Fujisawa} demonstrated that the Berry curvature $\Omega^z$ of the Cr$_{1+\delta}$Te$_2$ thin films can switch from positive to negative as $\delta$ varies, and that $\sigma_{xy}^{int}$ also changes its sign, crossing through zero.

\par
In conclusion, we observe skew scattering dominated AHE in the layered chalcogenide Cr$_{1+\delta}$Te$_2$ with $\delta \sim$ 0.33. The electronic band structure in the presence of SOC is quite fascinating with the presence of both positive and negative Berry curvature values in the momentum space, which together make the intrinsic AHE arising from band topology negligible. Additionally, we observe a node at the K point of the BZ in the unoccupied conduction band. It is worth performing suitable electron doping in the sample, which might shift the $E_F$ above the node, and one can observe a significant intrinsic anomalous Hall effect.  

\section*{Acknowledgments}
P.C. gratefully acknowledges the DST-INSPIRE program [Grant No. DST/INSPIRE Fellowship/2019/IF190532] for research assistance. M.N. extends his gratitude to CSIR, India, for research support [File No. 09/080(1131)/2019-EMR-I]. Access to the X-ray facilities in the Materials Characterisation Laboratory at the ISIS Facility is gratefully acknowledged. Part of this research was conducted at PETRA III, DESY, Germany, and we would like to thank Ann-Christin Dippel, Oleh Ivashko, and Fernando Igoa for their assistance with the P21.1 beamline (Proposal No. I-20221017). We also acknowledge the Department of Science and Technology (India) for financial support for the experimental work at PETRA III under the India@DESY collaboration program. J.S thanks S. N. Bose National Centre for Basic Science for Bridge fellowship.

\bibliographystyle{apsrev4-2} 
\bibliography{Citation}

\end{document}